\documentclass{article}
\usepackage{spconf,amsmath,graphicx}

\usepackage{booktabs}
\usepackage{multirow}

\title{IMPROVE BILINGUAL TTS USING DYNAMIC LANGUAGE AND PHONOLOGY EMBEDDING}
\name{Fengyu Yang, Jian Luan, Yujun Wang}
\address{Xiaomi AI Lab, Beijing, China}
\begin{document}
%\ninept
%
\maketitle
\begin{abstract}
%任务 传统方法 改进 结果
  %Non-native pronunciations of English result from the common linguistic phenomenon in which non-native users of any language tend to carry the intonation, phonological processes and pronunciation rules from their first language or first languages into their English speech. They may also create innovative pronunciations for English sounds not found in the speaker's first language.
  In most cases, bilingual TTS needs to handle three types of input scripts: first language only, second language only, and second language embedded in the first language. In the latter two situations, the pronunciation and intonation of the second language are usually quite different due to the influence of the first language. Therefore, it is a big challenge to accurately model the pronunciation and intonation of the second language in different contexts without mutual interference. This paper builds a Mandarin-English TTS system to acquire more standard spoken English speech from a monolingual Chinese speaker. We introduce phonology embedding to capture the English differences between different phonology. Embedding mask is applied to language embedding for distinguishing information between different languages and to phonology embedding for focusing on English expression. We specially design an embedding strength modulator to capture the dynamic strength of language and phonology. Experiments show that our approach can produce significantly more natural and standard spoken English speech of the monolingual Chinese speaker. From analysis, we find that suitable phonology control contributes to better performance in different scenarios.
\end{abstract}

\section{Introduction}

Nowadays, a bilingual text-to-speech(TTS) system is necessary for many application scenarios like voice assistant. For example, the names of English songs and movies are often directly embedded in Chinese responses. A straightforward way to build a bilingual TTS system is by collecting speech data from bilingual speakers. \cite{qian2009cross} proposed a shared hidden Markov model (HMM)-based bilingual TTS system, using a Mandarin-English corpus recorded by a bilingual speaker. \cite{fan2016speaker} presented a TTS system using a speaker and language factorized deep neural network(DNN) with a corpus of three bilingual speakers. However, mixed-lingual corpora are scarce while a large number of monolingual corpora are easily accessible.

Another way is to leverage monolingual speech data from different speakers \cite{latorre2006new, himawan2020speaker, latorre2005polyglot, zen2012statistical, sitaram2016experiments, xie2016kl}. \cite{latorre2005polyglot} proposes a polyglot synthesis method adapting the shared HMM states to the target speaker, trained on monolingual corpora. \cite{zen2012statistical} proposes to factorize speaker and language based on an HMM-based parametric TTS system. \cite{sitaram2016experiments} utilizes a combined phonetic space in two languages to build a code-switched TTS system based on HMM. \cite{xie2016kl} maps the senones between two monolingual corpora in two languages with a speaker-independent DNN ASR output based on HMM TTS. %\cite{he2012turning} uses a trajectory tiling approach to do cross-lingual speech synthesis on a monolingual speaker’s speech. \cite{sitaram2016experiments} utilizes a combined phonetic space in two languages to build an code-switched TTS system based on HMM. \cite{xie2016kl} maps the senones between two monolingual corpora in two languages with a speaker-independent DNN ASR output based on HMM TTS.

End-to-end TTS systems also extend to multilingual tasks using monolingual speech\cite{cao2019end, xue2019building, cao2020code, zhao2020towards, li2019bytes, nachmani2019unsupervised, zhang2019learning, liu2020tone}. \cite{li2019bytes} used Unicode bytes as a unified new language representation for multilingual TTS. 125 hours of speech were used and their system can read code-switching text, despite the problem of speaker inconsistency when cross-language. \cite{nachmani2019unsupervised} trained with designed loss terms preserving the speaker’s identity in multiple languages based on the VoiceLoop architecture \cite{taigman2017voiceloop}. The trained speech is recorded by 410 monolingual speakers speech from English, Spanish and German. \cite{zhang2019learning} used an adversarial loss term to disentangle speaker identity from the speech content, which trained with 550 hours of speech from 92 monolingual speakers. Limited by corpus size, \cite{liu2020tone} proposed tone embedding and tone classifier for tone preservation to generate utterances in a proper prosodic accent of the target language.

Generally, each speaker speaks only one language, leading to speaker and language characteristics being highly correlated. Using only monolingual corpora for bilingual or multilingual TTS easily leads to heavy accent carry-over in synthesized speech or inconsistent voice between languages. Actually, bilingual corpus helps deal with the problem. \cite{maiti2020generating} trained a TTS system transforming speaker embedding between languages from a bilingual speaker to other monolingual speakers for a high degree of naturalness. In this paper, we expect to utilize scarce bilingual corpora to acquire more standard spoken English from a monolingual speaker, which is highly correlated with phonology learning.

For example, in mixed-lingual utterances, the pronunciation of English by a non-native speaker, like Chinese, is strongly influenced by their native language and is most often different from the standard English pronunciation\cite{lee2018learning}. Mandarin derives pronunciation directly from the spellings of the word with different tones, which have a high grapheme-to-phoneme(g2p) correlation. In contrast, English is an alphabetic and highly non-phonemic language. 
In Consequence, native phonemic language speakers, whose pronunciation is influenced by the spelling of the word, often pronounce English words differently from standard English speakers\cite{baby2021non}. In mixed-lingual utterances, these speakers, despite qualified bilingual speakers, generally replace some English phonemes with the closest phoneme in their native language, resulting in mispronunciation and differences in phonology like articulation change and intonation variation\cite{he2012turning}.

Given these challenges, building a state-of-the-art bilingual TTS system requires special designs handling the English differences in phonology between mixed-lingual and monolingual utterances. In this paper, our contributions include: (1) introducing phonology embedding to capture the English differences between mixed-lingual and monolingual utterances; (2) proposing embedding mask to language embedding for distinguishing information between different languages and phonology embedding for focusing on expression between different phonology of English; (3) designing embedding strength modulator(ESM) to capture the dynamic information of language and phonology, which helps to generate more standard spoken English speech; (4) experiments showing that static and dynamic components in ESM can control different attributes of phonology. Phonology decomposition and control can make a contribution to more standard spoken English expression and better performance in different scenarios.

\section{Model Structure}
\label{sec:ms}

\begin{figure}[t]
  \centering
  \includegraphics[width=0.95\linewidth]{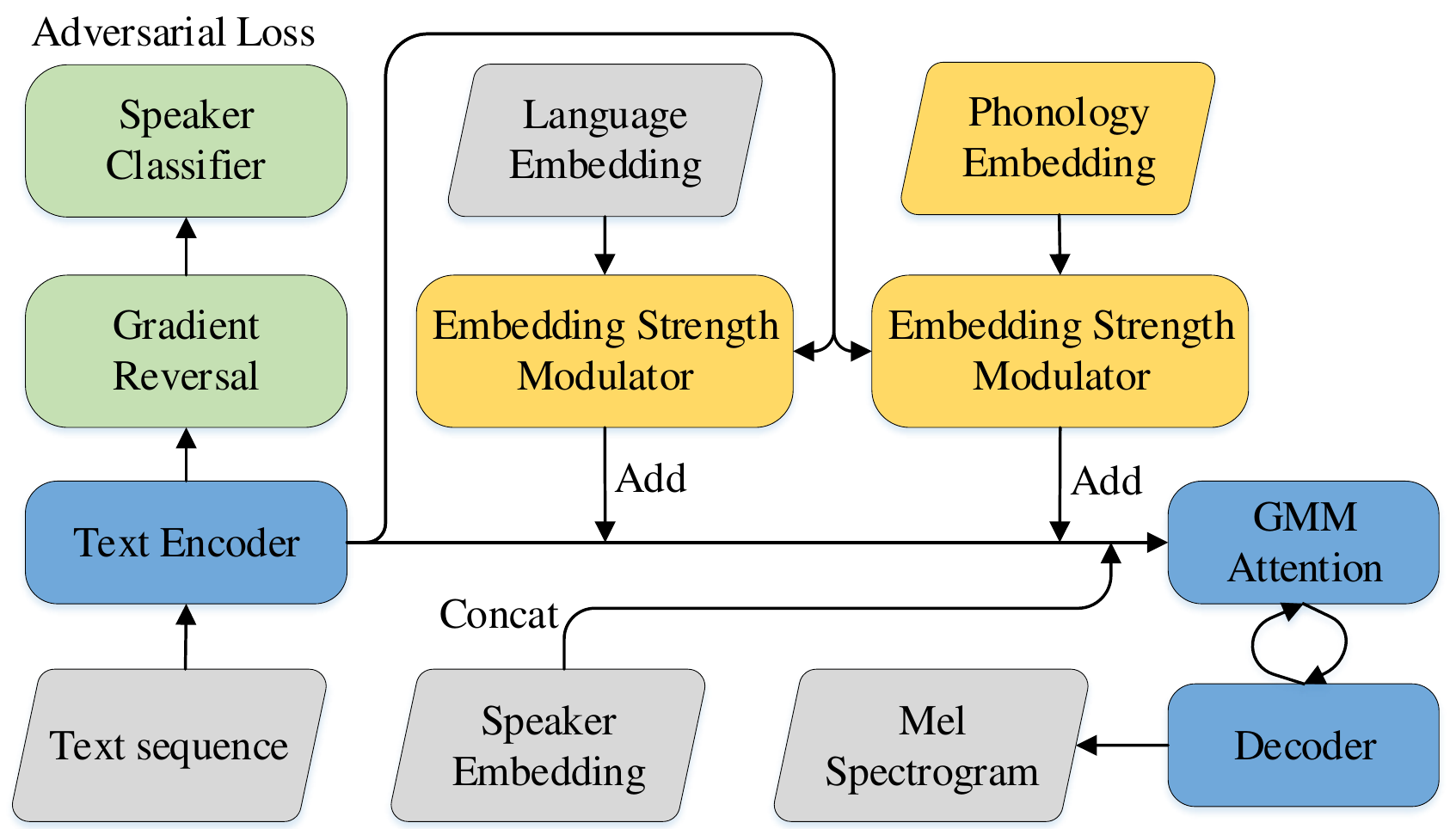}
  \caption{Overview of the proposed bilingual architecture with specially designed modules marked in yellow color.}
  \vspace{-0.5cm}
  \label{fig:tacotron2}
\end{figure}

Fig.~\ref{fig:tacotron2} illustrates the proposed bilingual TTS architecture. The encoder-attention-decoder backbone with speaker and language embedding will be described in Sec.~\ref{sec:bl}. The phonology embedding and specially designed masks for language and phonology embedding respectively will be described in Sec.~\ref{sec:lp}. The embedding strength modulator will be described in Sec.~\ref{sec:dn}.

\vspace{-0.3cm}
\subsection{Baseline}
\label{sec:bl}

%\begin{figure}[b]
%  \centering
%  \includegraphics[width=\linewidth]{Tacotron.pdf}
%  \caption{Overview of the baseline bilingual architecture.}
%  \label{fig:tacotron}
%\end{figure}

%As shown in Fig~\ref{fig:tacotron}, 
Our baseline system adopted from \cite{shen2018natural} is a popular Tacotron2\cite{shen2018natural}-based multilingual TTS architecture. It uses attention to bridge encoder and decoder. Language and speaker information are embedded in separate look-up tables. They are combined with the encoder output to distinguish different languages and speakers. Besides, an adversarially-trained speaker classifier is employed to disentangle text encoder output from speaker information. Mel-Lpcnet adopted from \cite{valin2019lpcnet} is used as a vocoder to reconstruct waveform from given mel-spectrogram.

The architecture takes phoneme sequences as inputs for both English and Mandarin. Their phoneme sets are simply concatenated and no phoneme is shared across. Tone or stress tokens are inserted into the phoneme sequence at the end of each syllable. For Mandarin, there are 4 lexicon tones and one neutral tone. Instead, there are 4 stress types for English includes the sentence, primary, secondary, and none. Moreover, prosodic break tokens are inserted into the input sequence as well. Finally, the expanded phoneme set contains: 73 Mandarin phonemes, 39 English phonemes, 5 Mandarin tones, 4 English stresses, Mandarin character boundary, English syllable boundary, English liaison symbol and 4 shared prosodic break types, i.e. prosodic word (PW), prosodic phrase (PPH), intonation phrase (IPH) and silence at the beginning or end.

\vspace{-0.2cm}
\subsection{Embedding mask}
\label{sec:lp}

\begin{figure}[t]
  \centering
  \includegraphics[width=6.8cm]{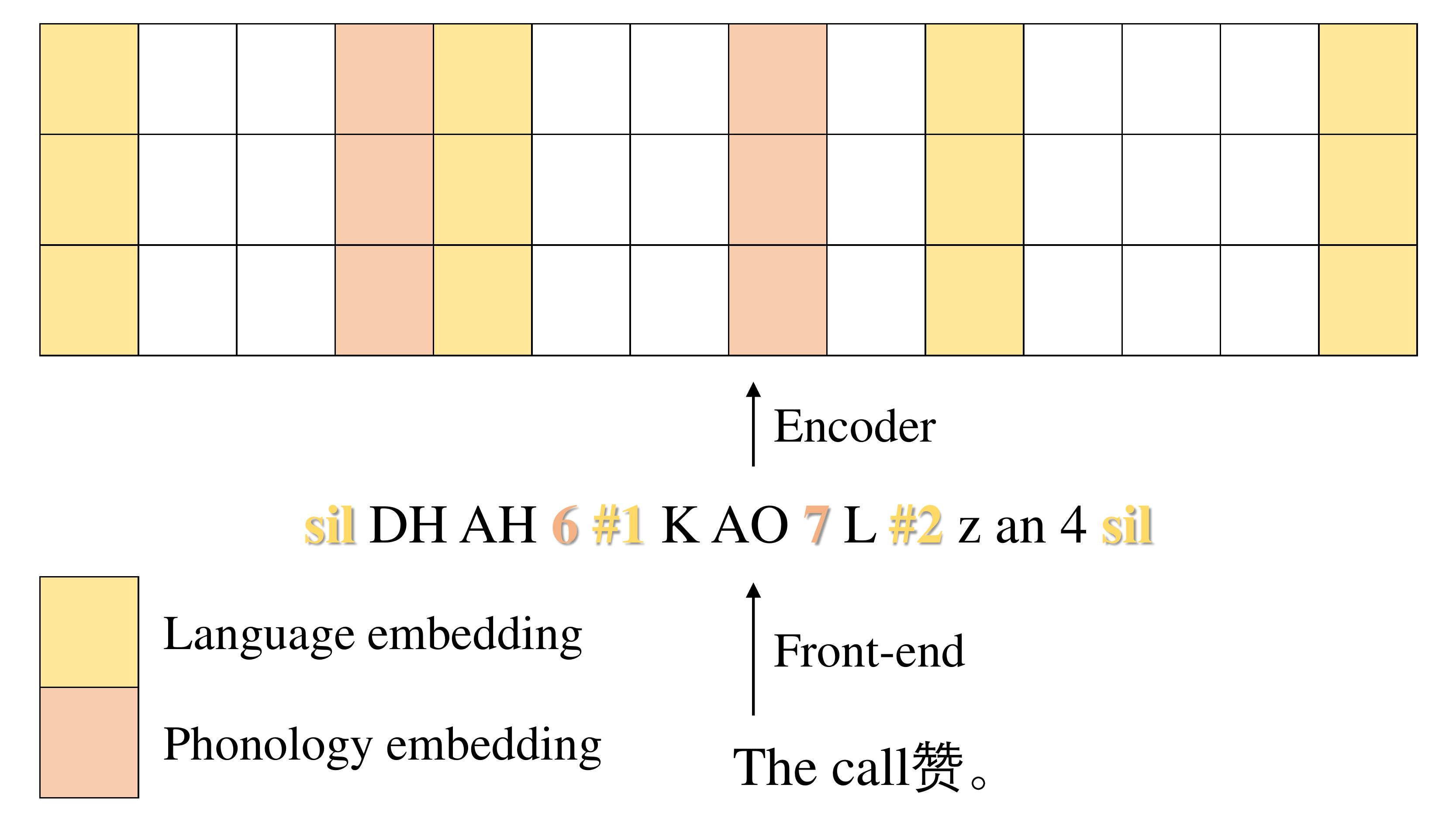}
  \caption{An illustration of how to mask embedding. Language and phonology embedding only applied to the highlighted position of encoder outputs. The symbols $\#$1, $\#$2, $\#$3 and /sil/ denote 4 shared prosodic break types. The numbers 1-5 denote tones of the previous Chinese syllable. The numbers 6-9 denote stresses of the previous English syllable.}
  \vspace{-0.5cm}
  \label{fig:embedding}
\end{figure}

Fig.~\ref{fig:embedding} shows an example of embedding mask in language and phonology embedding.

Instead of broadcasting language embedding to all the tokens of the input sequence, the proposed method applies language embedding only to the token types shared across languages, i.e. PW, PPH, IPH and /sil/. Because other token types are language-specific already and need no additional information to distinguish language.

On the other hand, to capture the English differences between the mixed-lingual and monolingual utterances, a special phonology embedding is designed. To focus on English expression, it is applied to all English-specific tokens, including 4 types of stresses, syllable boundary and liaison symbol.

\vspace{-0.05cm}
\subsection{Embedding strength modulator}
\label{sec:dn}

%With specially designed language and accent embeddings, the architecture can capture language-dependent and accent-related information, which does not consider the varying importance of them in each phoneme. Assuming the variety of the phoneme-level embeddings may have different strength and different contribution to the naturalness and fluency of the synthesized cross-lingual speech, we utilize a attention-based network supplying to catch the information associated to language and accent which has different strength fluctuating. To constrain the variety of phoneme-level embeddings, one static vector is applied with dynamic scalar weights, which we used as dynamic component. The original static embedding extracted the average and stationary information from language and accent is used as static component.
Even though the language and phonology embedding have been limited to only part of input tokens by masks, we think their strength should vary for different contexts. To capture the dynamic strength of languages and phonology, we propose an attention-based embedding strength modulator, whose framework is similar to \cite{xiong2020layer}.

\begin{figure*}[t]
  \centering
  \includegraphics[width=0.6\linewidth]{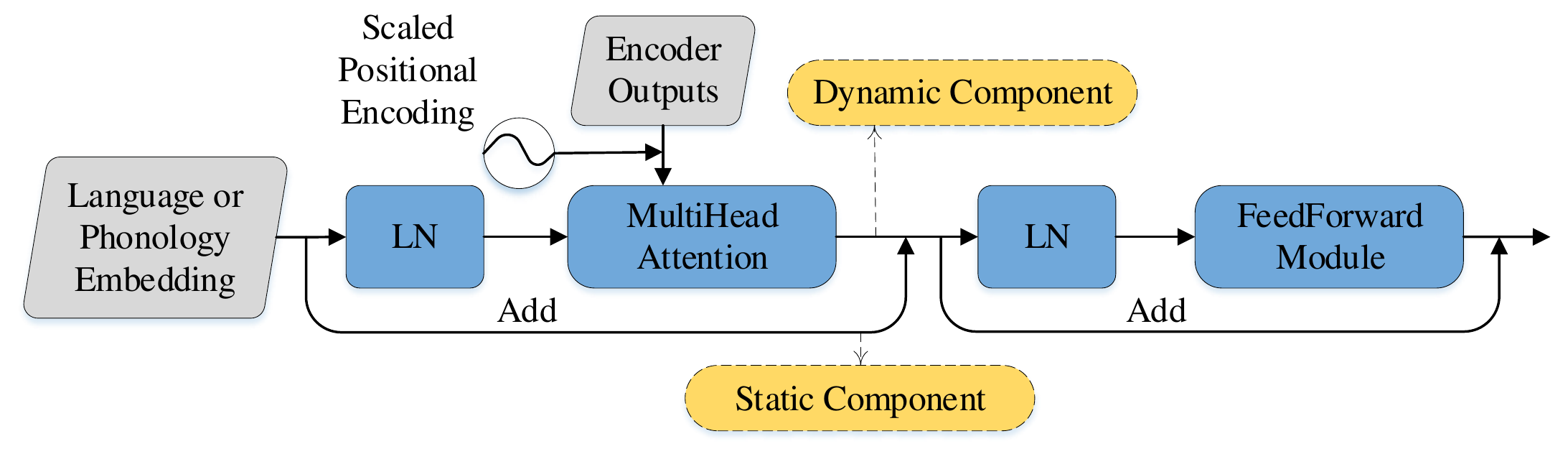}
  \caption{The structure of the embedding strength modulator of both language and phonology embedding.}
  \vspace{-0.3cm}
  \label{fig:decomposition}
\end{figure*}

The structure of the ESM is shown in Fig.~\ref{fig:decomposition}. There are two sub-networks in ESM: multi-head attention and a feed-forward network. The layer normalization and residual connection are applied to both of the sub-networks. Formally, from the encoder output with scaled positional encoding $E_o$, and the language or phonology embedding $LP$, the first sub-network $M_o$ and the second sub-network $F_o$ are calculated as:
\begin{equation}
  M_o={\rm MH}(E_o,\ {\rm LN}(LP),\ {\rm LN}(LP))+LP,
  \label{eq:M}
\end{equation}
\begin{equation}
  F_o={\rm FFN}({\rm LN}(M_o))+M_o.
  \label{eq:F}
\end{equation}
where MH(query, key, value), FFN(·) and LN(·) are multi-head attention, feed-forward network and layer normalization respectively. Since the attention key and value ($LP$) have only one item, the energy need not be normalized by softmax operation. Instead, each head in multi-head attention is computed by:
\begin{equation}
  head_h={\alpha}_h{\cdot}V_h=\frac{Q_h{\cdot}K_h}{\Vert Q_h\Vert\Vert K_h\Vert}{\cdot}V_h,
  \label{eq:head}
\end{equation}
where $\Vert \cdot \Vert$ is the L2 norm of the last dimension, $\{Q, K, V\}$ represent query, key and value through linear transformation respectively and the strength ${\alpha}$ is a scaled cosine similarity between the query and key to be in the range of [-1, 1].

In particular, there are two components in Fig.~\ref{fig:decomposition} marked in yellow color. The original embedding learned for each language and phonology is regarded as a static component. While the output of multi-head attention, the static embedding multiplied by a dynamic weight, is regarded as a dynamic component. We will analyze the roles each component of language and phonology embedding play in Sec.~\ref{ana}.

\vspace{-0.2cm}
\section{Experiments}
\label{sec:ex}

\vspace{-0.2cm}
\subsection{Basic setups}
Models are trained with proprietary datasets composing three kinds of high-quality speech: (1) bilingual corpus from two Chinese speakers, 45000 and 25000 Mandarin utterances for female and male speakers respectively, 9000 mixed-lingual utterances and 9000 English utterances for both speakers; (2) English corpus from two American speakers, 9000 and 25000 English utterances for female and male speakers respectively; (3) Mandarin corpus from a female Chinese speaker, 9000 Mandarin utterances for cross-lingual experiments.

Labels of the above corpora in language and phonology embedding are described in Tab.~\ref{tab:label}. Mandarin utterances are all from Chinese speakers. Their language is labeled Mandarin and no English phonology label is required. For plain English utterances, the corpora recorded by both American and Chinese speakers are labeled as English for language and Standard-English for phonology. These Chinese speakers pronounce English well and the corpora recording by American speakers are used as supplementary English datasets and are beneficial to speaker learning. Particularly, since English parts in mixed-lingual utterances are in a small amount and are mostly words and abbreviations with heavy Chinese phonology, we label them Mandarin for language and Chinese-English for phonology, treated as Mandarin utterances.

\begin{table*}[t]
  \caption{Labels of trained corpora in language and phonology embedding.}
  \label{tab:label}
  \centering
  \begin{tabular}{lcc|cc}
    \toprule
    \multicolumn{3}{c}{\textbf{Corpus}} & \textbf{\rule{0pt}{9pt}Language embedding} & \textbf{Phonology embedding} \\
    \midrule
    \multirow{3}{*}{(1) Chinese speaker} & \multirow{3}{*}{Train} & \rule{0pt}{9pt}Mandarin & Mandarin & None \\
    \multirow{3}{*}{} & \multirow{3}{*}{} & \rule{0pt}{9pt}Mixed-lingual & Mandarin & Chinese-English \\
    \multirow{3}{*}{} & \multirow{3}{*}{} & \rule{0pt}{9pt}English & English & Standard-English \\
    \hline
    (2) American speaker & Train & \rule{0pt}{12pt}English & English & Standard-English \\
    \hline
    (3) Chinese speaker & Test & \rule{0pt}{12pt}Mandarin & Mandarin & None\\
    \bottomrule
    \vspace{-0.5cm}
  \end{tabular}
\end{table*}

The additional inputs of the learned speaker (64-dim), language and phonology embedding (both 512-dim same with the dimensions of encoder output) are injected into the backbone. In ESM, the first sub-network includes 8-head multi-head attention and the feed-forward sub-network consists of two convolution networks with 2048 and 512 hidden units. Linguistic inputs have been introduced in Sec.~\ref{sec:lp} and for acoustic features, we use an 80-band mel-spectrogram extracted from 16kHz waveforms. We built the following systems for comparison:
\begin{itemize}
\item {\bf BASE}: Baseline system with senteneial language embedding as described in Sec.~\ref{sec:bl};
\item {\bf EM}: Baseline system with specially designed language and phonology embedding as described in Sec.~\ref{sec:lp};
\item {\bf ESM}: Baseline system with specially designed language and phonology embedding through ESM as described in Sec.~\ref{sec:dn}.
\end{itemize}

\subsection{Subjective evaluation}

We conduct Mean Opinion Score (MOS) evaluations of speech naturalness and speaker similarity via subjective listening tests. 20 speakers are asked to listen to the generated 20 English utterances and 10 mixed-lingual utterances. MOS results are reported in Tab.~\ref{tab:mos}. Except for parts of samples in listing tests, generated Mandarin demos of this monolingual speaker are also shown in demo pages\footnote{Samples can be found from: https://fyyang1996.github.io/esm/}.

\begin{table}[t]
  \caption{The MOS of different systems with confidence intervals of 95\%.}
  \label{tab:mos}
  \centering
  \begin{tabular}{lccc}
    \toprule
    \textbf{Model} & \textbf{BASE} & \textbf{EM} & \textbf{ESM} \\
    \midrule
    $Naturalness$ & 3.81$\pm$0.12  & 4.03$\pm$0.10 & \textbf{4.39$\pm$0.08}~~~             \\
    $Similarity$  & 3.79$\pm$0.12  & 3.91$\pm$0.11 & \textbf{4.04$\pm$0.10}~~~             \\
    \bottomrule
  \end{tabular}
\end{table}

We can find that the EM system with masked embedding brings better performance on both speech naturalness and speaker similarity than the conventional BASE system. It indicates that masked embedding captures features that better represent language and phonology. For the further proposed embedding strength modulator, we find that by capturing the dynamic strength of language and phonology system ESM achieves significantly better performance than the EM system. It demonstrates that the dynamic strength of language and phonology is beneficial to speech naturalness and speaker similarity of generated speech. %But without sufficient bilingual and monolingual corpora recorded by different speakers, the scores of speaker similarity in the three systems are all not too high.

\vspace{-0.2cm}
\subsection{ESM component analysis}
\label{ana}

\begin{figure}[t]
\begin{minipage}[b]{0.44\linewidth}
  \centering
  \centerline{\includegraphics[width=4.5cm]{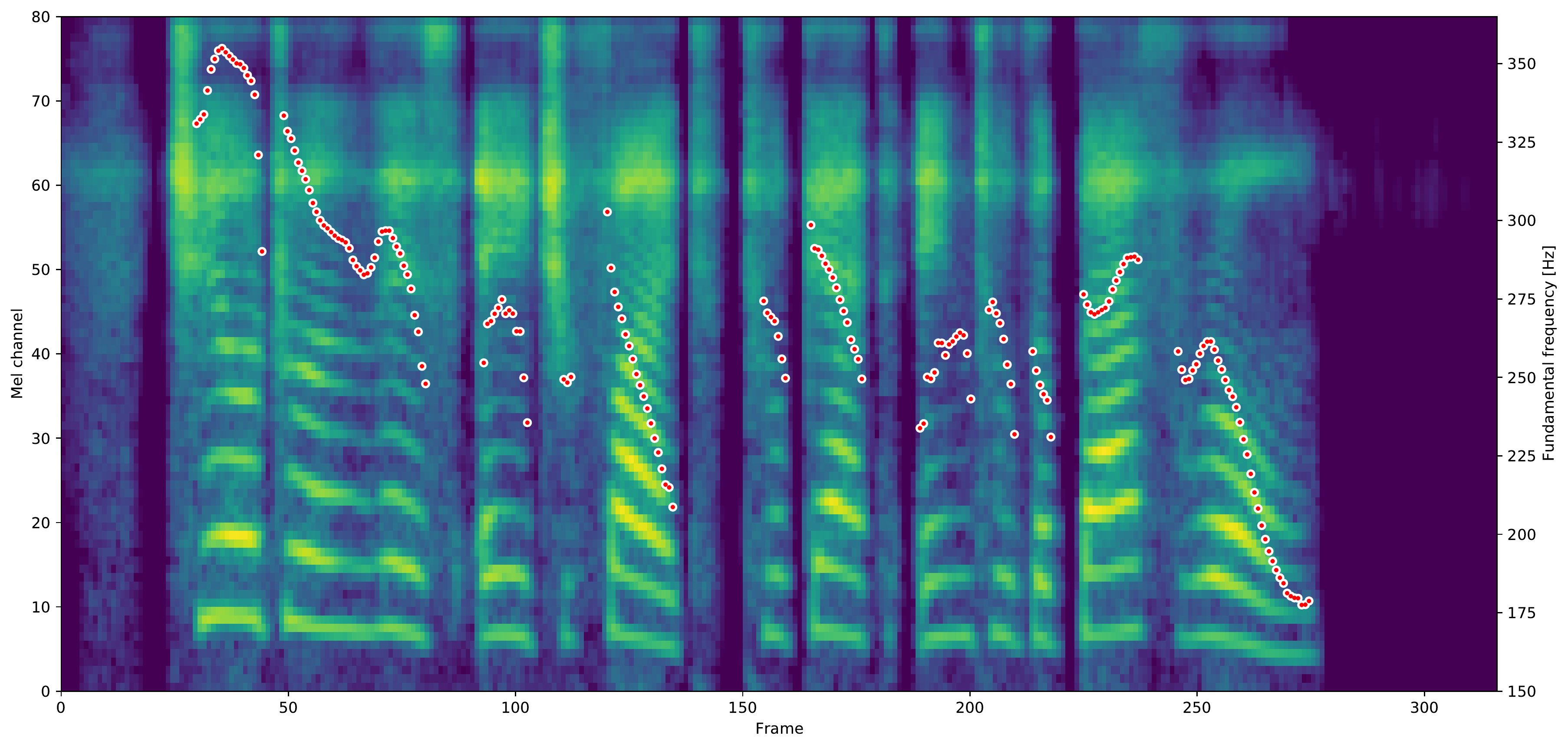}}
%  \vspace{2.0cm}
  \centerline{(a) Base combination}\medskip
\end{minipage}
\hfill
\begin{minipage}[b]{0.44\linewidth}
  \centering
  \centerline{\includegraphics[width=4.5cm]{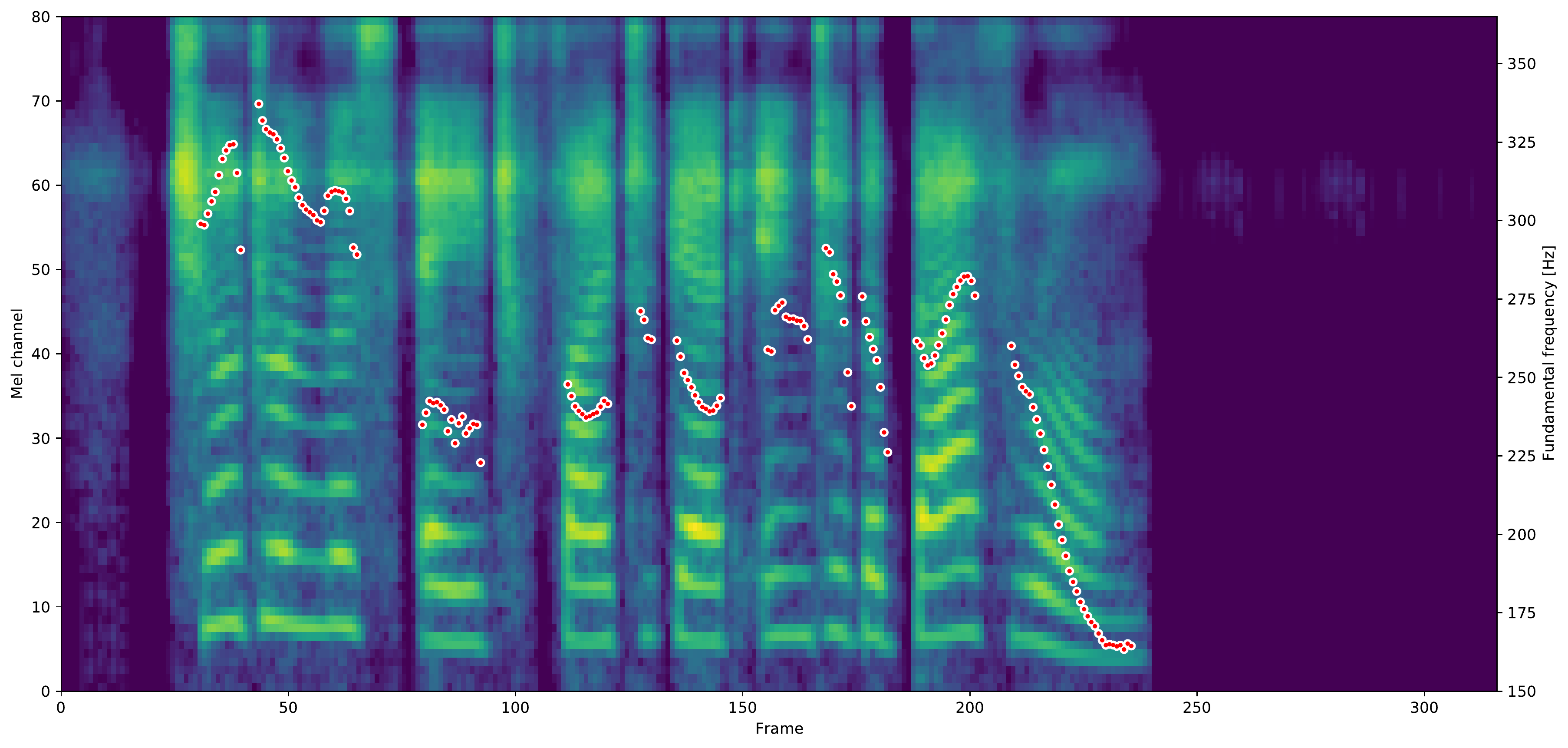}}
%  \vspace{1.5cm}
  \centerline{(b) Reference combination}\medskip
\end{minipage}
\begin{minipage}[b]{0.44\linewidth}
  \centering
  \centerline{\includegraphics[width=4.5cm]{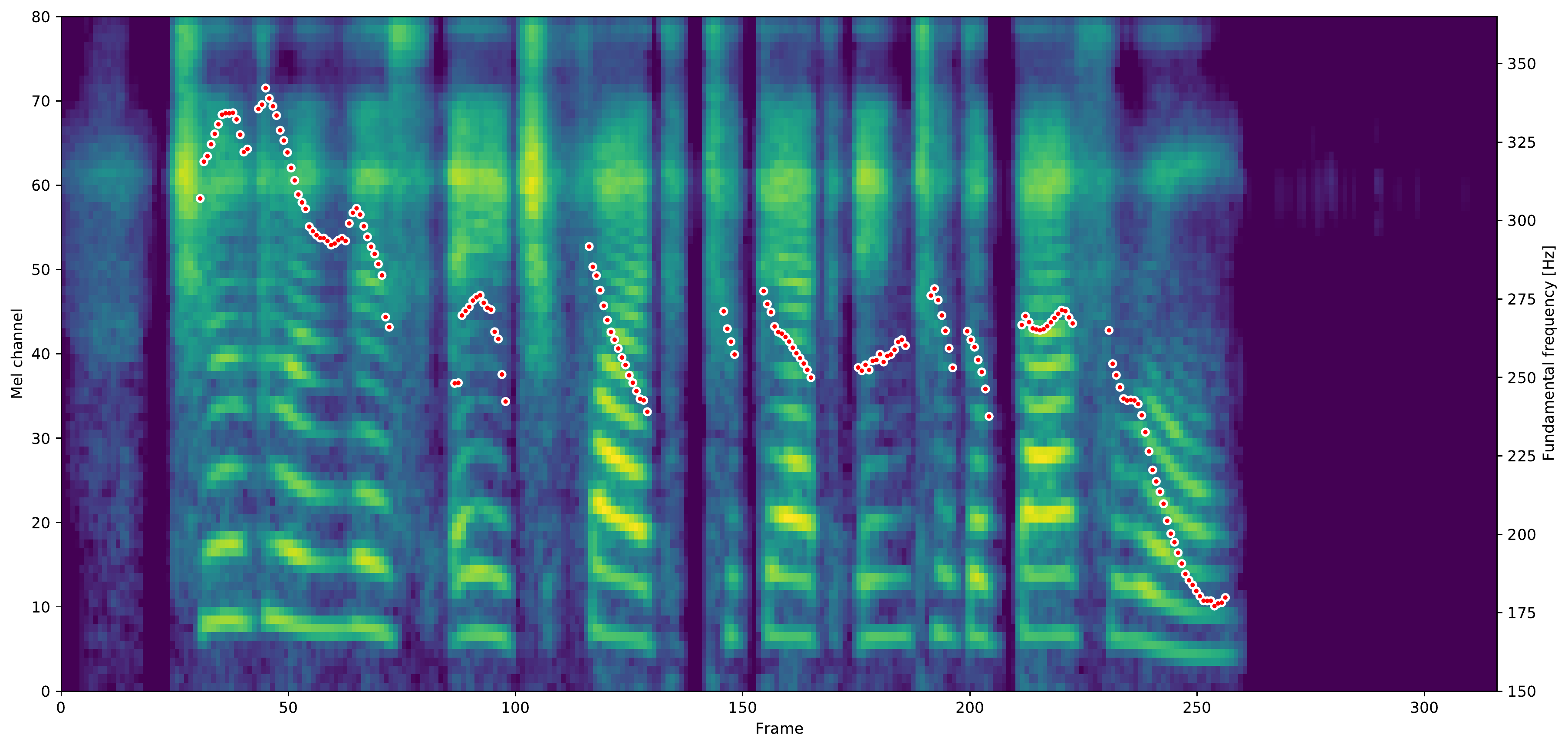}}
%  \vspace{1.5cm}
  \centerline{(c) Dynamic phonology embedding}\medskip
\end{minipage}
\hfill
\begin{minipage}[b]{0.44\linewidth}
  \centering
  \centerline{\includegraphics[width=4.5cm]{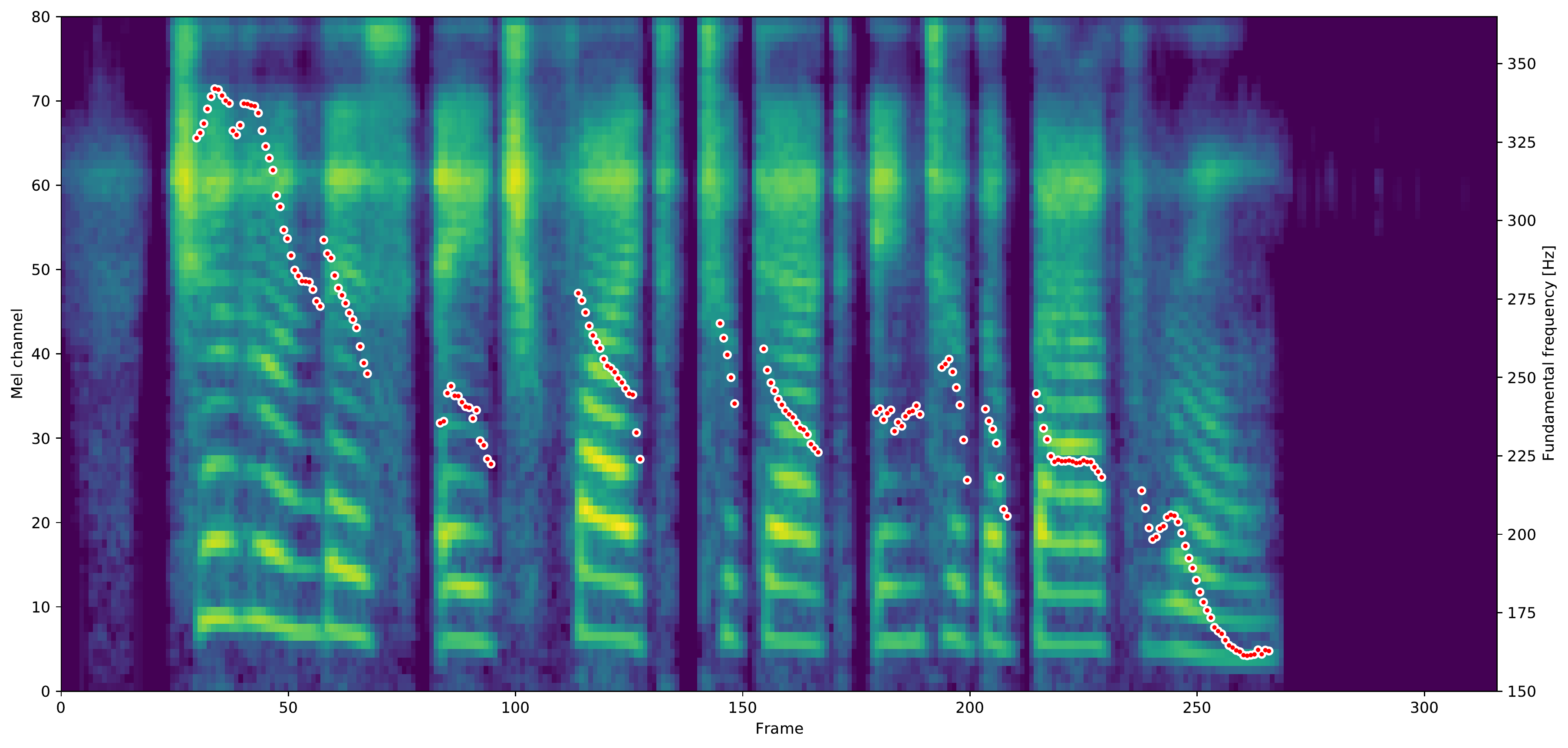}}
%  \vspace{1.5cm}
  \centerline{(d) Static phonology embedding}\medskip
\end{minipage}
\begin{minipage}[b]{0.44\linewidth}
  \centering
  \centerline{\includegraphics[width=4.5cm]{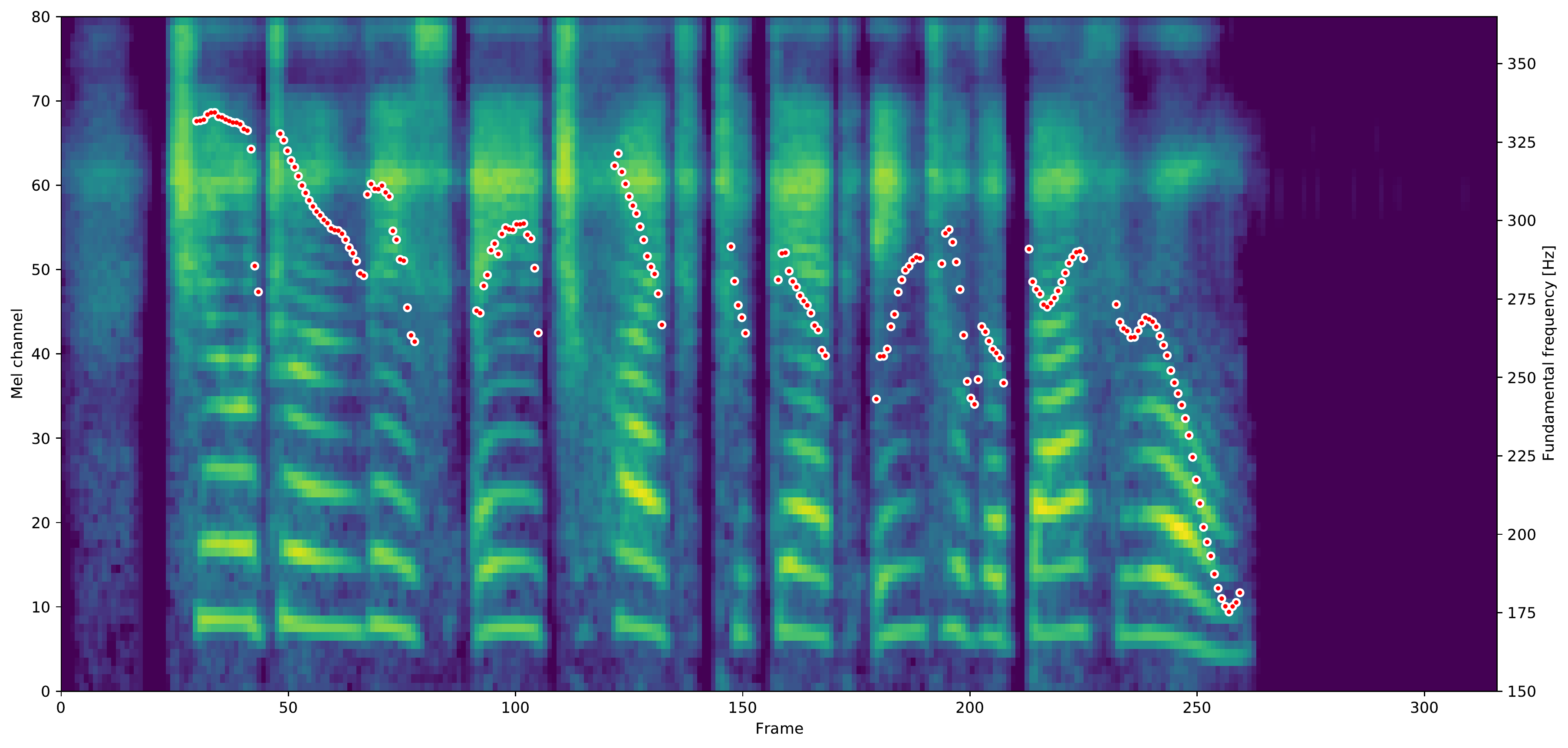}}
%  \vspace{1.5cm}
  \centerline{(e) Dynamic language embedding}\medskip
\end{minipage}
\hfill
\begin{minipage}[b]{0.44\linewidth}
  \centering
  \centerline{\includegraphics[width=4.5cm]{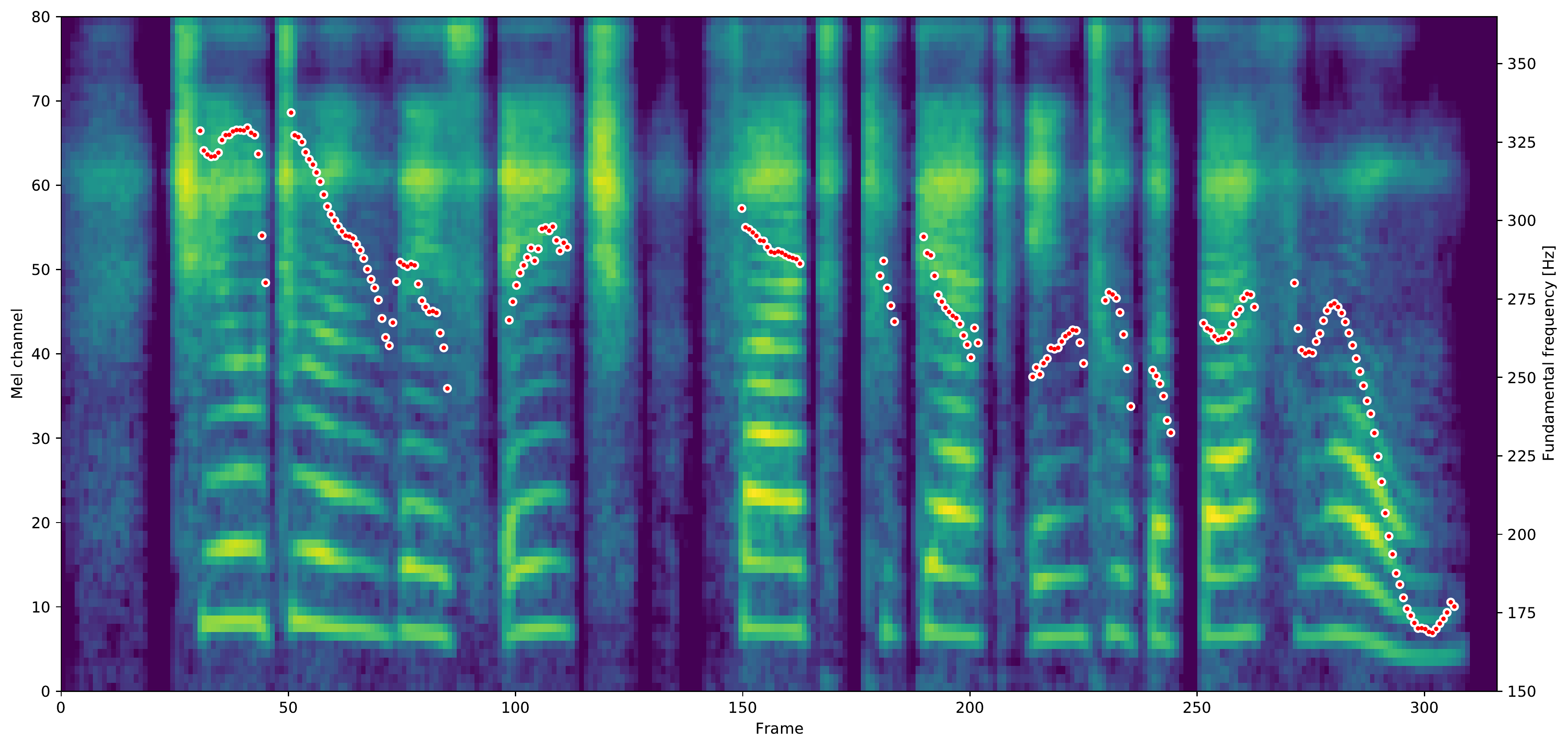}}
%  \vspace{1.5cm}
  \centerline{(f) Static language embedding}\medskip
\end{minipage}
\caption{Spectrogram and F0 of a test sentence generated by different combinations, which refers to 2.1 in demo page.}
\vspace{-0.5cm}
\label{fig:mel_f0}
\end{figure}

As mentioned above, the output of ESM may be regarded as the combination of a static component and a dynamic component. One simple method of analyzing the contribution of each component is to condition the model on only one component at each time. In the generation phase, we replace the static or dynamic component from Mandarin label to English label for language embedding or from Chinese-English phonology label to Standard-English phonology label for phonology embedding respectively. Fig.~\ref{fig:mel_f0} shows the spectrogram and F0 contour, extracted by parselmouth\cite{jadoul2018introducing}, of the same sentence synthesized with six kinds of label combinations as described below:
\begin{itemize}
\item[(a)] Base combination: using Mandarin and Chinese-English phonology labels both in dynamic and static components;
\item[(b)] Reference combination: using English and Standard-English phonology labels both in dynamic and static components;
\item[(c)] Based on (a), replacing dynamic phonology embedding from Chinese-English to Standard-English phonology.
\item[(d)] Based on (a), replacing static phonology embedding from Chinese-English to Standard-English phonology;
\item[(e)] Based on (a), replacing dynamic language embedding from Mandarin to English;
\item[(f)] Based on (a), replacing static language embedding from Mandarin to English;
\end{itemize}

Empirically, we find that each component represents articulation, intonation, speaking rate and pause duration changes respectively, which influence phonology collectively. Listing to the samples of (a) and (c) in the demo page, we can easily hear about articulation changes between them, which is difficult to be caught sight of. Perceptually, the trend of F0 values in Fig.~\ref{fig:mel_f0}(d) is different from that in Fig.~\ref{fig:mel_f0}(a), showing that static phonology embedding major affects intonation. Fig.~\ref{fig:mel_f0}(e) shows that replacing the dynamic language embedding from Mandarin to English causes a gradual compression of the spectrogram and F0 values in the time domain. We believe that the dynamic language embedding encodes the information correlated with speaking rate variation. Besides, syllables in Fig.~\ref{fig:mel_f0}(f) have distinct intervals compared with that in Fig.~\ref{fig:mel_f0}(a), which demonstrates that static language embedding represents the average duration of pauses. More demos and be found in the demo page.
%Perceptually, compared with Fig.~\ref{fig:mel_f0}(a), the peak value of F0 in Fig.~\ref{fig:mel_f0}(c) has changed within a phoneme in several positions, which illustrates dynamic accent embedding captured the information of phonological change mostly. 

%s , height=2.5cm

\subsection{Control}

%\textbf{Enhance expressiveness } To enhance the expressiveness of a plain English text, we double the dynamic components of both language and phonology embedding while remaining their static components. The "double" herein means that the final vector has a double distance of the reference vector to the base vector. For language embedding, the reference is English and the base is Mandarin. While for phonology embedding, the reference is Standard English and the base is Chinese-English. Fig.~\ref{fig:pitch} shows the F0 trajectories of a sentence with base, reference and doubled embeddings. It can be easily found that the many syllables in the base have a large pitch range, appearing obvious Mandarin style. The reference seems to be English style while the doubled is more fluent and has a larger pitch range than the reference.

%\begin{figure}[t]
%  \centering
%  \includegraphics[width=8.8cm]{000004_pitch.pdf}
%  \caption{F0 values of a test utterance generated by different combinations. Audios can be found in 3.1 of the demo page.}
%  \label{fig:pitch}
%\end{figure}

To validate the above analysis, we conduct MOS evaluations of speech naturalness and speaker similarity via subjective listening tests. 20 speakers are asked to listen to the generated 15 English utterances for enhancing English expressiveness and 15 mixed-lingual utterances for smooth mixed-lingual transition. Demos can be found in 3 and 4 on the demo pages.

\begin{figure}[t]
  \centering
  \includegraphics[width=1\linewidth]{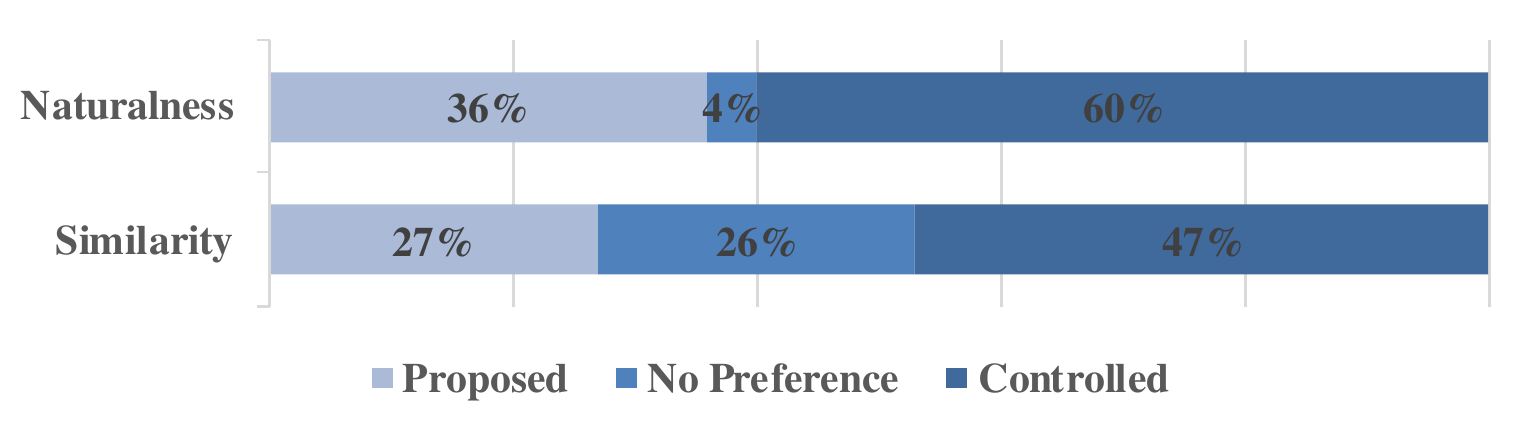}
  \vspace{-0.8cm}
  \caption{A/B preference results for control in enhancing expressiveness or not with confidence intervals of 95\% and p-value\textless0.0001 from the t-test.}
  \vspace{-0.1cm}
  \label{fig:abtest-eng}
\end{figure}

\textbf{Enhance expressiveness } To enhance the expressiveness of a plain English text, we double the dynamic components of both language and phonology embedding while remaining their static components. The "double" herein means that the final vector has a double distance of the reference vector to the base vector. For language embedding, the reference is English and the base is Mandarin. While for phonology embedding, the reference is Standard English and the base is Chinese-English. Fig.~\ref{fig:abtest-eng} shows the results of MOS evaluations. We find that by the "double" operation herein system ESM achieves significantly better performance than the ESM system on speech naturalness. It indicates that the control operation enhances English expressiveness significantly.

\begin{figure}[t]
  \centering
  \includegraphics[width=0.98\linewidth]{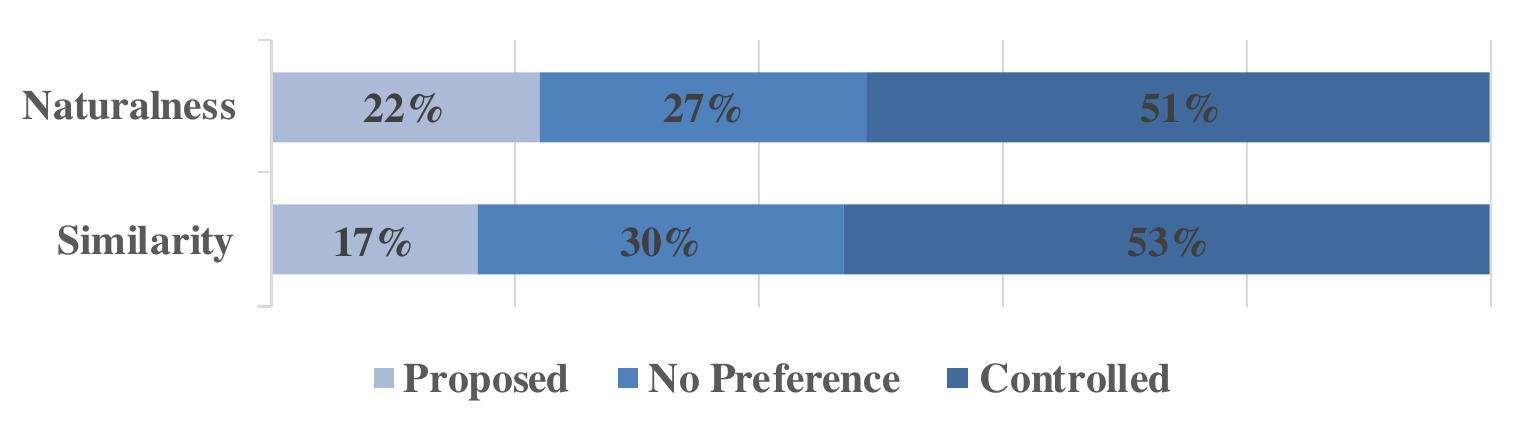}
  \vspace{-0.5cm}
  \caption{A/B preference results for control in smooth transition or not with confidence intervals of 95\% and p-value\textless0.0001 from the t-test.}
  \vspace{-0.3cm}
  \label{fig:abtest-mix}
\end{figure}

%\noindent\textbf{\rule{0pt}{13pt}Smooth transition }
\textbf{Smooth transition } When synthesizing a mixed-lingual text, we modify the language labels of embedded English words from Mandarin to English while phonology labels of them from Chinese-English to Standard-English. Particularly, their static component of phonology embedding remains Chinese-English. In this way, the English words will have standard-English articulation but more compatible intonation with the context of Chinese words. Fig.~\ref{fig:abtest-mix} shows the results of MOS evaluations. It can be found that the controlled ESM system brings better performance on both speech naturalness and speaker similarity than the proposed ESM system.  It demonstrates that the control operation is beneficial to smooth mixed-lingual transition.

  \vspace{-0.15cm}
\section{Conclusions}
\label{sec:co}

This paper builds a Mandarin-English TTS system for a monolingual Chinese speaker. We introduce phonology embedding and a special designed mask for language and phonology embedding. They are employed to distinguish two languages and the English phonological differences between monolingual and embedded cases respectively. Furthermore, the proposed embedding strength modulator enables language and phonology embedding to be variable with token context. Experiments show that our approach can produce significantly more natural and standard spoken English speech than baseline. Ablation analysis on different components demonstrates that English phonology can be tuned effectively for various scenarios.

\bibliographystyle{IEEEbib}
\bibliography{strings,refs}

\end{document}